\documentclass[letterpaper, twocolumn,superscriptaddress, showpacs,preprintnumbers,amsmath,amssymb] {revtex4}
\usepackage{graphicx}
\usepackage{dcolumn}
\usepackage{bm}
\usepackage{hyperref}
\usepackage{subfigure}
\usepackage{verbatim}
\hypersetup{
    colorlinks,
    citecolor=red,
    filecolor=blue,
    linkcolor=blue,
    urlcolor=blue
}

\begin{document}

\title{Three `species' of Schr\"odinger cat states in an infinite-range spin model}

\author{Bo Zhao}
\affiliation{Physics Department, Princeton University, Princeton, NJ 08544, USA}

\author{Merritt C. Kerridge}
\affiliation{Physics Department, Princeton University, Princeton, NJ 08544, USA}

\author{David A. Huse}
\affiliation{Physics Department, Princeton University, Princeton, NJ 08544, USA}

\begin{abstract}
We explore a transverse-field Ising model that exhibits both spontaneous symmetry-breaking and eigenstate thermalization.
Within its ferromagnetic phase, the exact eigenstates of the Hamiltonian of any large but finite-sized system are all Schr\"odinger cat states: superpositions of states with `up' and `down' spontaneous magnetization.  This model exhibits two dynamical phase transitions {\it within} its ferromagnetic phase:  In the lowest-temperature phase the magnetization can macroscopically oscillate between up and down.  The relaxation of the magnetization is always overdamped in the remainder of the ferromagnetic phase, which is divided in to phases where the system thermally activates itself {\it over} the barrier between the up and down states, and where it quantum tunnels.
\end{abstract}


\maketitle

\section{Introduction}

The dynamical properties of isolated many-body quantum systems have long been of interest, due to their role in the fundamentals of quantum statistical mechanics.  More recently, experiments approximating this ideal of isolated many-body quantum systems have become feasible in systems of trapped atoms \cite{weiss,bloch} and ions \cite{blatt}, and as a consequence this topic has received renewed attention.  It appears that a broad class of such systems obey the Eigenstate Thermalization Hypothesis (ETH) \cite{deutsch,sred,tasaki,rigol,review}.  The ETH asserts that each exact many-body eigenstate of a system's Hamiltonian is, all by itself, a proper microcanonical ensemble in the thermodynamic limit, in which any small subsystem is thermally equilibrated, with the remainder of the system acting as a reservoir.  In the present paper we present some interesting results for an infinite-range transverse-field Ising model that obeys the ETH and also has spontaneous symmetry-breaking.

Quantum many-body systems with static randomness may fail to obey the ETH due to many-body Anderson localization stopping thermalization \cite{pwa,baa,pal}.  The interesting interplay of many-body localization and discrete symmetry-breaking was recently explored in Refs. \cite{huse,pea,va,jonas}; the present paper instead explores an example of the interplay of the ETH and Ising symmetry breaking.

We start with the infinite-range transverse-field Ising model:
\begin{equation}\label{eq:model}
H_0=-\frac{1}{N}\sum_{1=i<j}^{N}s_{i}^{z}s_{j}^{z}-\Gamma\sum_{i=1}^{N}s_{i}^{x} ~,
\end{equation}
where $\vec s_i=(s_i^x,s_i^y,s_i^z)=\vec\sigma_i/2$, and $\vec\sigma_i$ are the Pauli operators for the spin-1/2 at `site' $i$.  We choose to set $\hbar=k_B=1$.
This model has been extensively studied recently, particularly as an example for exploring quantum information issues where it is known as the `Lipkin-Meshkov-Glick model'; see, e.g., \cite{wilms} and references therein.
One can determine and use many properties of the exact eigenstates of this Hamiltonian, but there are extensive degeneracies in its spectrum due to its symmetry under any permutation of the $N$ spins, which give it eigenstates that do not obey the ETH.  Thus we add static random Ising interactions to the Hamiltonian to break the permutation symmetry and lift all the degeneracies (with probability one), so the full Hamiltonian of the system we consider is
\begin{equation}\label{eq:ham}
H=H_0+H_1=H_0+\frac{\lambda}{N^p}\sum_{1=i<j}^{N}\varepsilon_{ij}s_{i}^{z}s_{j}^{z} ~,
\end{equation}
where $1\geq\lambda>0$, the $\varepsilon_{ij}$ are independent Gaussian random numbers of mean zero and variance one, and the power $p$ satisfies 
$1/2<p<1$.  As we argue below, the eigenstates of this Hamiltonian should obey the ETH, but the randomness is weak enough so that many of the 
properties of $H_0$, such as the thermodynamics, are unchanged and can be used in our analysis.  This randomness is too weak to produce any 
localization.  We have chosen to put the randomness on the interactions, since we have found in exact diagonalizations that this produces much 
better thermalization at numerically accessible system sizes as compared to, e.g., only making the local transverse fields random.

\begin{figure}[!htbp]
\begin{center}
\includegraphics[width = \columnwidth]{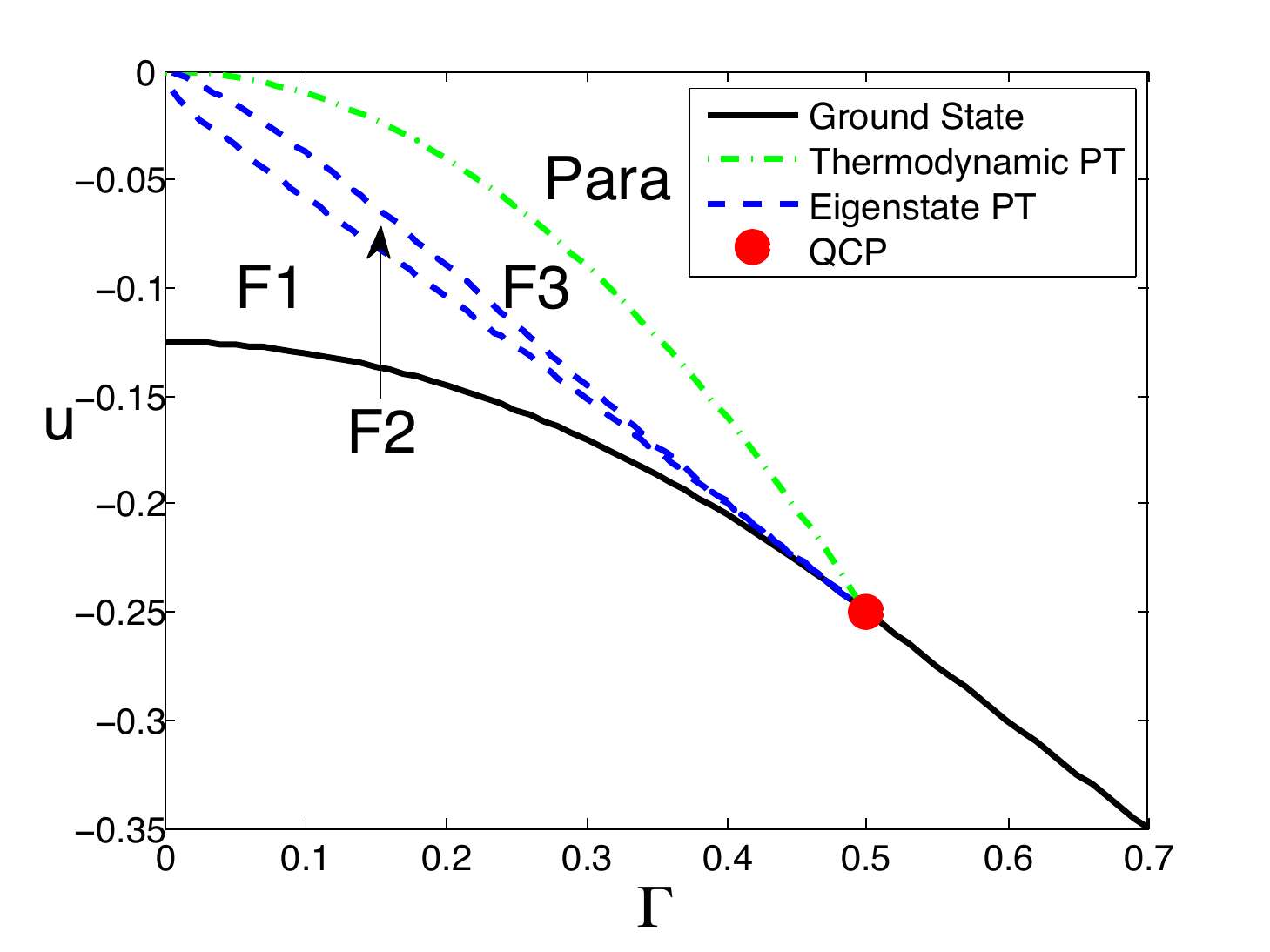}
\caption{(color online) The phase diagram of our model.  $u$ is the energy per spin and $\Gamma$ is the transverse field. The ground state (zero temperature) is indicated by the solid (black) line, with the quantum critical point (QCP) indicated.  The dot-dashed (green) line is the thermodynamic phase transition (PT) between the paramagnetic and ferromagnetic (F) phases.  There are two dynamical (eigenstate) phase transitions within the ferromagnetic phases, indicated by the dashed (blue) lines.  See the text for discussions of the sharp distinctions between these three phases F1, F2 and F3.}
\label{fig:phase diagram}
\end{center}
\end{figure}

We now briefly summarize our results, before deriving and discussing them in more detail below.  The phase diagram of this spin model as a function of the energy per spin $u=\langle H\rangle/N$ and the transverse field $\Gamma$ is shown in Fig. 1.  There are the usual two thermodynamic phases of a ferromagnet: the paramagnetic phase (Para) at high energy and/or high $|\Gamma|$, and the ferromagnetic phase (F) when both $|\Gamma|$ and the energy are low enough.  In the ferromagnetic phase, for any finite $N$, we can ask about the dynamics of the system's order parameter.  There are three regimes of behavior that are sharply distinguished from one another in the thermodynamic limit $N\rightarrow\infty$:  At the highest energies within region F3 of the ferromagnetic phase the system is a sufficiently large thermal reservoir for itself so that the most probable path by which it flips from `up' to `down' magnetization under unitary time evolution is by thermally activating itself over the free energy barrier between the two ordered states.  At lower energies (F1 and F2) the barrier is higher and wider and as a result the reservoir is inadequate, so the system quantum tunnels through the barrier when it flips the Ising order parameter.  At the lowest energies in region F1 one can in principle prepare a state that is a linear combination of two Schr\"odinger cat eigenstates of $H$ that will coherently oscillate via macroscopic quantum tunneling between up and down magnetizations.  In the intermediate energy regime (F2) the magnetization dynamics due to quantum tunneling is always overdamped.

Throughout the ferromagnetic phase, the exact eigenstates of $H$ for any finite $N$ are Schr\"odinger cat states that are superpositions of up and down magnetized states, and the properties of these cats differ in the three regimes of the ferromagnetic phase that are indicated in Fig. 1.  Thus the two phase transitions between these three dynamically distinct ferromagnetic phases are not only dynamical phase transitions but also `eigenstate phase transitions' \cite{huse}, while the equilibrium thermodynamic properties are perfectly analytic through these two phase transitions.

We consider the infinite-range model not only because this allows a controlled calculation of this novel physics within the ferromagnetic phases, but also because finite-range, finite-dimensional models obeying ETH do not show these features.  In the latter models the free energy needed to flip the magnetization by making a domain wall and sweeping it across the system is sub-extensive, while at any nonzero temperature the system is a reservoir of extensive size, so a large system will always flip via the thermal process without macroscopic quantum tunneling through the energy barrier; phases F1 and F2 thus do not exist for such models.  One can also consider intermediate cases of transverse-field Ising models with interactions that fall off as a power of the distance between spins.  When this power is small enough, the resulting free energy barrier to flip the magnetization in the ferromagnetic phase is extensive, and we thus expect phases F1 and F2 to also occur in those models, although we do not see a way to simply calculate the locations of the phase boundaries as we can for the infinite-range model.

\section{Thermodynamics}

First we examine the the unperturbed Hamiltonian $H_0$, which has the same thermodynamics as our full model $H$.  The Hamiltonian $H_0$ commutes with all permutations of the spins, as does the total spin operator:
\begin{equation}
\vec{S}\equiv\sum_{i=1}^{N}\vec{s_{i}} ~.
\end{equation}
The magnitude $S^2$ of the total spin squared also commutes with $H_0$, so we can choose a set of eigenstates of $H_0$ that are also eigenstates of $S^2$.  This unperturbed Hamiltonian only depends on the total spin and thus can be written as
\begin{equation}
H_0=-\frac{1}{2N}S_{z}^{2}-\Gamma{}S_{x}+\frac{1}{8} ~.
\end{equation}
The magnitude $S$ of the total spin is of order $N$ and ranges from zero up to $S=N/2$.  For the thermodynamics (but not the dynamics) of this system in the limit of
large $N$ we can treat the components of $\vec S$ classically and ignore their nonzero commutators when obtaining the extensive
thermodynamic properties (energies, entropies, magnetizations).  
The ground state of $H_0$ always has the maximum value of $S=N/2$.  For $\Gamma\geq 1/2$, the ground state is paramagnetic with the spins polarized along the $x$-direction: $S_z/N=0$, $S_x/N=1/2$ and $u=\langle H_0\rangle/N=-\Gamma/2$.  For $|\Gamma|<1/2$, the two nearly-degenerate ground states are ferromagnetic, with $S_z/N=\pm(1/2)\sqrt{1-4\Gamma^2}$, $S_x/N=\Gamma$ and $u=-(1+4\Gamma^2)/8$.

For each value of $S$ the spectrum of $H_0$ has $(2S+1)$ eigenenergies.  These eigenenergies and the corresponding eigenstates can be approximated for large $S$ using a discrete version of the WKB method \cite{braun}, as we discuss below.  Each energy level in the spectrum of $H_0$ with total spin $S/N=\bar{s}<1/2$ has entropy per spin
\begin{equation}\label{eq:entropy}
\Sigma(\bar{s})=-\big[(\frac{1}{2}-\bar{s})\ln(\frac{1}{2}-\bar{s})+
(\frac{1}{2}+\bar{s})\ln(\frac{1}{2}+\bar{s})\big] ~,
\end{equation}
due to all the different ways one can add together $N$ spin-1/2's to get total spin $S$.  

Since we are interested here in eigenstates, which are at a given energy $U=Nu$, we will do the statistical mechanics in the microcanonical ensemble.  For a given transverse field $\Gamma$ and energy $U$, the equilibrium (most probable) state of the system is the one that maximizes the entropy, which means minimizing the total spin $S$.  To minimize $S$ for a given $U$, clearly we set $S_y=0$, since $S_y$ does not appear in the Hamiltonian.  In the paramagnetic phase the total spin points along the $x$-direction and the equilibrium value of the total spin is thus $S_{eq}/N=|u/\Gamma|$.  In the ferromagnetic phase the system can go to higher entropy (lower total spin) for a given $U$ by making $S_z/N\neq 0$.  Some algebra shows that in the ferromagnetic phase, which is $|\Gamma|<1/2$ and $-(1+4\Gamma^2)/8\leq u<-\Gamma^2$, the equilibrium is at $S_x/N=\Gamma$, $S_z/N=\pm \sqrt{2(-u-\Gamma^2)}$ and $S_{eq}/N=\sqrt{-2u-\Gamma^2}$.  The line of critical points separating the para- and ferromagnetic phases is $u=-\Gamma^2$, for $|\Gamma|<1/2$, as indicated in Fig. 1; this critical line ends at the quantum critical points at $|\Gamma|=1/2$, $u=-1/4$.

\section{Eigenstate thermalization}

Due to its full symmetry under all permutations of the $N$ spins, the Hamiltonian $H_0$ is integrable, with all the good quantum numbers associated with this permutation symmetry, including the magnitude $S$ of the total spin.  We want to study a more generic system, so we add to the Hamiltonian the small term $H_1$ (see Eq. (2)) to break the permutation symmetry, lift all degeneracies, and make the eigenstates thermal.  The only symmetry that remains in our full $H$ is the Ising ($Z_2$) symmetry under a global rotation of all spins by angle $\pi$ about their $x$ axes.

For a given $U$ and $\Gamma$, the eigenstates of $H_0$ have total spin ranging from the minimum and equilibrium value $S_{eq}$ up to the maximum value of $S=N/2$.  To make these in to thermal eigenstates we need $H_1$ to perturb the system enough so that the eigenstates of the full $H$ are linear combinations of all these total spin values, weighted as at thermal equilibrium.  At first order, the perturbation we are adding, $H_1$, flips at most two spins, so it can change the total spin by at most $\pm 2$.  The spectrum of $H_0$ at each value of total spin $S$ contains $(2S+1)$ eigenenergies spread over a range of energy that is of order $S$.  Thus the level-spacing in the spectrum of $H_0$ at a given $S$ remains of order one in the limit of large $N$.  This is reflected in the dynamics under $H_0$, which is spin precession about the mean field, and the mean field is of order one, so the rate of precession is also of order one.

For the eigenstates of $H$ to strongly and thermally mix the different values of $S$ we thus need the matrix elements of $H_1$ between states at different $S$ to be large compared to the (order-one) level spacing of $H_0$ in the large $N$ limit.  This is why we require that the exponent $p$ in the definition of $H_1$ satisfies $p<1$, since this is the condition for these matrix elements to diverge in the large $N$ limit.  This should be sufficient to make all the eigenstates of $H$ thermal in that limit.  We have not yet found a way to actually {\it prove} that this is sufficient to make all the eigenstates of $H$ satisfy the ETH, but below we provide some numerical evidence for this from exact diagonalization of finite-size systems.

\section{Dynamics}

In addition to making sure that $H_1$ is strong enough to thermalize the system, we also want it to be weak enough so that we can use the well-understood dynamics of $H_0$ in our analysis.  By restricting the exponent $p$ to be greater than $1/2$, in the large $N$ limit the effective field that each spin is precessing about is the mean field from $H_0$, with only a small correction from $H_1$ that vanishes as $N\rightarrow\infty$.  This small correction is enough to thermalize the system for $1/2<p<1$, which is the range where the perturbation due to $H_1$ on a single spin's dynamics vanishes for $N\rightarrow\infty$, while the perturbation to the dynamics of the full many-body system diverges.  In this regime, the system's primary dynamics is the $S$-conserving dynamics due to $H_0$, which is spin precession at a rate of order one, and, assuming we are in the ferromagnetic phase, `attempts' at rate of order one to tunnel through the energy barrier between total $S_z$ up and down.  At a rate that is slower by a power of $N$, the dynamics due to $H_1$ allow `hopping' between different values of the total spin $S$, and thus thermalization to the equilibrium probability distribution of $S$ dictated by the entropy $\Sigma$.  And at a rate that is slower still, exponentially slow in $N$, the system succeeds in crossing the free energy barrier between up and down magnetizations.  It is the separation between these three time scales that allows us to systematically understand the dynamics of this system for large $N$.

Since our full Hamiltonian $H$ has Ising symmetry under a global spin flip, and the randomness in $H_1$ means there are no exact degeneracies in the spectrum of $H$, for finite $N$ any eigenstate of $H$ is either even or odd under this Ising symmetry (with probability one).  In the ferromagnetic phase, this means the exact eigenstates of $H$ are all Schr\"odinger cat states that are either even or odd linear combinations of states with total $S_z$ up and down.  These two equal and opposite values of $S_z$ are extensive, thus `macroscopically' different, which is why it is appropriate to call these eigenstates `Schr\"odinger cats'.

Next we examine the rate at which this system, in its ferromagnetic phase, will flip from the up state with positive total $S_z$ to the down state with negative $S_z$ under the unitary time evolution due to its Hamiltonian $H$.  There are two steps to this process: First the system gets `excited' from its usual (high entropy) total spin $S_{eq}$ `up' to a larger total spin $S$ with lower entropy, with a probability $\sim\exp{\{-N(\Sigma(S_{eq}/N)-\Sigma(S/N))\}}$ given by the resulting decrease of the entropy.  As $S$ is increased, the energy barrier, whose top is at energy $E=-\Gamma S$, decreases.  For $U\geq -\Gamma N/2$, one way the system can flip is to increase $S$ enough so that $U\geq -\Gamma S$ and then it will simply cross over the top of the barrier without quantum tunneling.  In the higher-energy part (F3) of the ferromagnetic phase, in the limit of large $N$ this is the dominant process that flips the magnetization: the system `thermally activates' itself (via its unitary time-evolution) to a low-entropy, high-total-spin state where the energy barrier can be crossed without quantum tunneling.  The `height' of the {\it entropy} barrier it must cross to do this is extensive:
\begin{equation}\label{eq:entropybarrier}
N\Delta\Sigma=N(\Sigma(\sqrt{-2u-\Gamma^2})-\Sigma(-u/\Gamma)) ~,
\end{equation}
where $\Delta\Sigma$ is the reduction in entropy per site needed to go over the barrier.

If the system does not or can not go over the energy barrier by increasing the total spin $S$, then in order to flip the magnetization it must quantum tunnel through the barrier.  For large $N$, this tunneling probability can be estimated using a version of the WKB method \cite{braun}.  We will summarize how this is done and its results, but not present all the details, which follow Ref. \cite{braun}.  The total $S_z$ serves as the `position', while the operator $-\Gamma S_x$ serves as the `kinetic energy'.  What we need to calculate is the probability of the system tunneling between positive and negative total $S_z$ for a given total spin $S=N\bar{s}$ satisfying $S_{eq}\leq S < -U/\Gamma$.  This probability behaves as $\sim\exp{(-N\gamma)}$, with $\gamma(u,\bar{s},\Gamma)$ being an intensive quantity.  If we define a scaled `position' $x=S_z/N$, the WKB `turning points' adjacent to the barrier are at $x=\pm x_t$ with
\begin{equation}
x_t=\sqrt{2}\cdot\sqrt{-u-\Gamma^2-\sqrt{\Gamma^2(2u+\bar{s}^2+\Gamma^2)}}\,.
\end{equation}
Then the intensive factor in the exponent of the tunneling probability is given by the WKB tunneling integral
\begin{equation}\label{eq:gamma}
\gamma = 2\int_{0}^{x_t}{\rm arcosh}{\frac{-u-\frac{1}{2}x^2}{\Gamma\sqrt{\bar{s}^2-x^2}}}\,dx\,.
\end{equation}

Since the probabilities of being `excited' to total spin $S$ and of quantum tunneling through the barrier with total spin $S$ are both exponentially small in $N$, in the limit of large $N$ the dominant process by which the magnetization flips is given by a standard `saddle point' condition.  The total spin $S=N\bar{s}$ at which the system tunnels is the value that maximizes the product of these two probabilities and thus minimizes the quantity
\begin{equation}\label{eq:saddlepoint}
\alpha(u,\Gamma)={\rm min}_{\bar{s}}\{\Sigma(\sqrt{-2u-\Gamma^2})-\Sigma(\bar{s})+\gamma(u,\bar{s},\Gamma)\} ~.
\end{equation}
We have located the saddle point numerically at many points within the ferromagnetic phase and it appears to always be unique, without any discontinuities as the parameters $u$ and $\Gamma$ are varied.  Some straightforward analysis shows that in the higher-energy part (F3) of the ferromagnetic phase where
\begin{equation}
\frac{\pi\Gamma}{\sqrt{-u-\Gamma^2}}\geq\ln\frac{\Gamma-2u}{\Gamma+2u}
\end{equation}
the saddle point is `thermal': the `entropy cost' of going to higher $S$ is less than the `tunneling cost', and the system goes over the barrier without any quantum tunneling.  We call the eigenstates in this regime `thermal cats', since these Schr\"odinger cat states flip by thermally activating themselves over the barrier.  In regions F1 and F2 this inequality is instead false, and the system quantum tunnels through the barrier at a value of $S$ satisfying $S_{eq}\leq S < -U/\Gamma$, so the eigenstates are instead `quantum cats'.  The location of the dynamical phase transition between phases F2 and F3 is given by converting the above inequality (10) to an equality.
Near the quantum critical point ($\Delta\Gamma=1/2-|\Gamma|\ll1$), this transition line becomes exponentially adjacent to (and above) the straight line $u=-|\Gamma|/2$: $u=-1/4+(\Delta\Gamma/2)+O(\exp(-1/\sqrt{\Delta\Gamma}))$; near $\Gamma=0$, it behaves as a power law: $-u\approx (|\Gamma|\sqrt{\pi/4})^{4/3}$.

Within the lower-energy phases (F1 and F2) of `quantum cats', there is a second dynamical phase transition within the ferromagnetic phase.  This is also a `spectral phase transition' \cite{huse} in the level-spacing statistics of the eigenenergies.  It one starts with a state (not an eigenstate) that is magnetized up, the rate at which the system crosses the barrier to down is $\sim\exp{(-N\alpha(u,\Gamma))}$, and as a result the uncertainty of the energy of this initial `up' state must be at least this large, by the time-energy uncertainty relation.  Compare this minimum energy uncertainty to the typical many-body level spacing $\sim\exp{(-N\Sigma(\sqrt{-2u-\Gamma^2}))}$ of the eigenstates of $H$ at that energy.  There is clearly a sharp change at the phase transition line, which is the line where $\alpha(u,\Gamma)=\Sigma(\sqrt{-2u-\Gamma^2})$.  This transition line between phases F1 and F2 is shown in Fig 1.  The location of this transition was obtained numerically, since we do not have a simple closed-form expression for $\alpha(u,\Gamma)$.
Near the quantum critical point ($\Delta\Gamma=1/2-|\Gamma|\ll1$), this transition line also becomes exponentially adjacent to (but now below) the straight line $u=-|\Gamma|/2$: $u=-1/4+(\Delta\Gamma/2)-O(\exp(-1/\sqrt{\Delta\Gamma}))$; near $\Gamma=0$, it is logarithmically tangent to the u-axis: $-u\sim{}1/(\ln|\Gamma|)^2$.

In the lowest-energy phase (F1) where
\begin{equation}
\alpha(u,\Gamma)>\Sigma(\sqrt{-2u-\Gamma^2}) ~,
\end{equation}
the many-body eigenstates come in almost-degenerate pairs that are well separated in energy from other pairs of eigenstates.  These eigenstates are still thermal in their fluctuations near equilibrium, so the inter-pair eigenenergy spacings have the level statistics of the Gaussian Orthogonal Ensemble (GOE).  Each pair of eigenstates within phase F1 is well-approximated by the two states
\begin{equation}
|n,\pm\rangle=\frac{(|\uparrow\rangle_n\pm |\downarrow\rangle_n)}{\sqrt{2}} ~,
\end{equation}
where $|\uparrow\rangle_n$ is a thermal state that is magnetized up, while $|\downarrow\rangle_n$ is its opposite under the global spin flip symmetry; here the label $n$ refers to such a pair of eigenstates.  In this regime (F1), one can make an up-magnetized initial state using a simple linear combination of only these two eigenstates, and this linear combination will then oscillate with a frequency $\sim\exp{(-N\alpha(u,\Gamma))}$.  Thus in this phase the Schr\"odinger cats can in principle be made to oscillate between `alive' (up) and `dead' (down).

In the higher-energy parts of the ferromagnetic phase (F2 and F3 in Fig. 1) where
\begin{equation}
\alpha(u,\Gamma)<\Sigma(\sqrt{-2u-\Gamma^2}) ~,
\end{equation}
the energy associated with the tunneling is large compared to the many-body level spacing, so there are no closely degenerate pairs of states from which we can make a coherently oscillating cat.  Instead, the eigenstates are of the form
\begin{equation}
|n\rangle=(|\uparrow\rangle_n + (-1)^{z_n} |\downarrow\rangle_n)/\sqrt{2} ~,
\end{equation}
with each eigenstate being either even ($z_n=0$) or odd ($z_n=1$) under the global spin flip; here the label $n$ refers to just one eigenstate.  States of opposite symmetry ($z_n \neq z_m$) that are nearly degenerate are not made out of the same thermal states in each well, so $_n\langle\uparrow|\uparrow\rangle_m\rightarrow\delta_{nm}$ for $N\rightarrow\infty$.  To make a state that is initially magnetized up requires coherently adding together exponentially many eigenstates in order to destructively cancel all the amplitudes for down magnetizations.  Under the unitary time evolution this special initial linear combination will dephase and the probability of down magnetization will increase, but presumably in an overdamped fashion as the state relaxes to equal probability of up and down magnetization.  Thus we expect the dynamics of the average magnetization to always be an overdamped relaxation in phases F2 and F3.  In phase F1, on the other hand, one can in principle prepare an initial state whose macroscopic magnetization will oscillate, as discussed above.  However, even in phase F1, if one starts in a generic state that has a given energy density and is magnetized up, this will also be a linear combination of exponentially many eigenstates and also presumable show an overdamped relaxation of the average magnetization.

\section{Numerical Evidence}

We now present some numerical results about the thermalization properties and the distinction between phases F1 and F3 (phase F2 is too narrow to clearly see in the size systems we can diagonalize).  We exactly diagonalize Hamiltonians with up to $N=15$ spins.
We first put the Hamiltonian into a block diagonal form using basis states that are even and odd with respect to the system's $Z_2$ Ising symmetry, and then diagonalize each sector numerically to obtain all of the eigenstates within that sector.
For finite $N$, the coefficient $\lambda$ 
in the disorder term $H_1$ in Eq. (2) matters.  
If $\lambda$ is too large the system will become a spin glass rather than a ferromagnet.
Thus $\lambda$ needs to be carefully chosen to be large enough for our finite systems to show thermalization, but small enough to avoid the spin glass regime.
After some exploration, we chose to use the parameters $\lambda=0.7$, $p=3/4$ and $\Gamma=1/8$ for our exact diagonalizations.

\begin{figure}[!htbp]
\begin{center}
\includegraphics[width = \columnwidth]{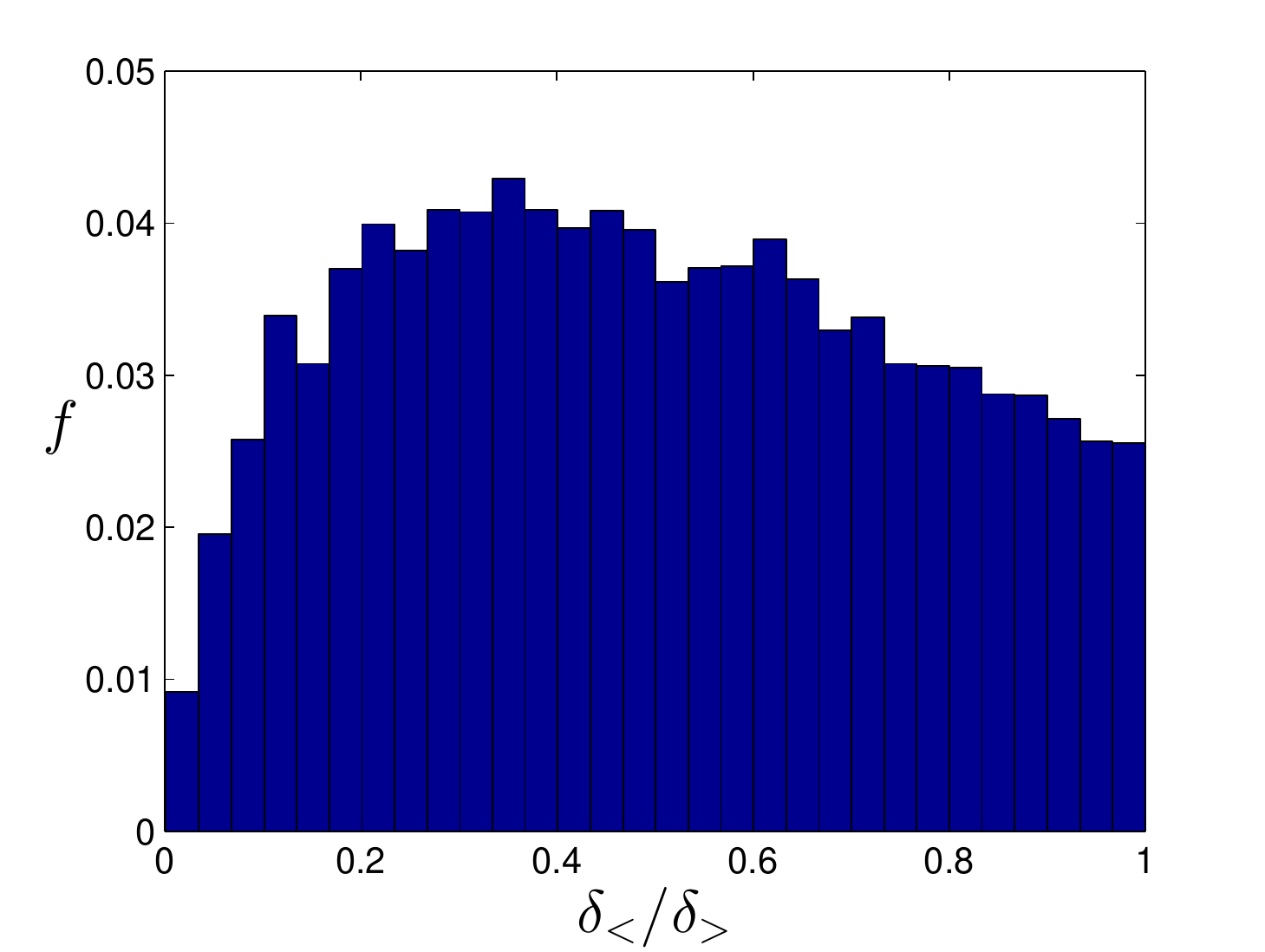}
\caption{(color online) The level-spacing statistics using 100 realizations of $H$ at $N=15$ in phase F1 within the even sector. $\delta_</\delta_>$ is the ratio between the smaller level spacing $\delta_<$ to the larger level spacing $\delta_>$ for three consecutive eigenenergies in the even sector. $f$ is the relative frequency for each bin in this histogram.}
\label{fig:GOE}
\end{center}
\end{figure}

First, we show the level-spacing statistics, which should be GOE if the eigenstates are thermal.  This must be done within one $Z_2$ symmetry sector, since there is no level-repulsion between states in different sectors.
We look at each set of three consecutive levels in one sector and denote $\delta_<$ as the smaller level spacing and $\delta_>$ as the larger level spacing. Then the histogram of the ratio $\delta_</\delta_>$ can be compared to GOE level statistics \cite{atas}.  The even sector results in the F1 phase for 100 realizations at $N=15$ are shown in Fig. 2.
We see the expected strong level repulsion, consistent with the thermalization.
All other phases and symmetry sectors were also examined and the results are also thermal, since phase F1 is the lowest-energy phase and
thus the most difficult to thermalize. 

\begin{figure}[!htbp]
  \subfigure[\, Eigenstate distance in phase F1]{
    \label{fig:F3} 
    \begin{minipage}[b]{\columnwidth}
      \centering
      \includegraphics[width = \textwidth]{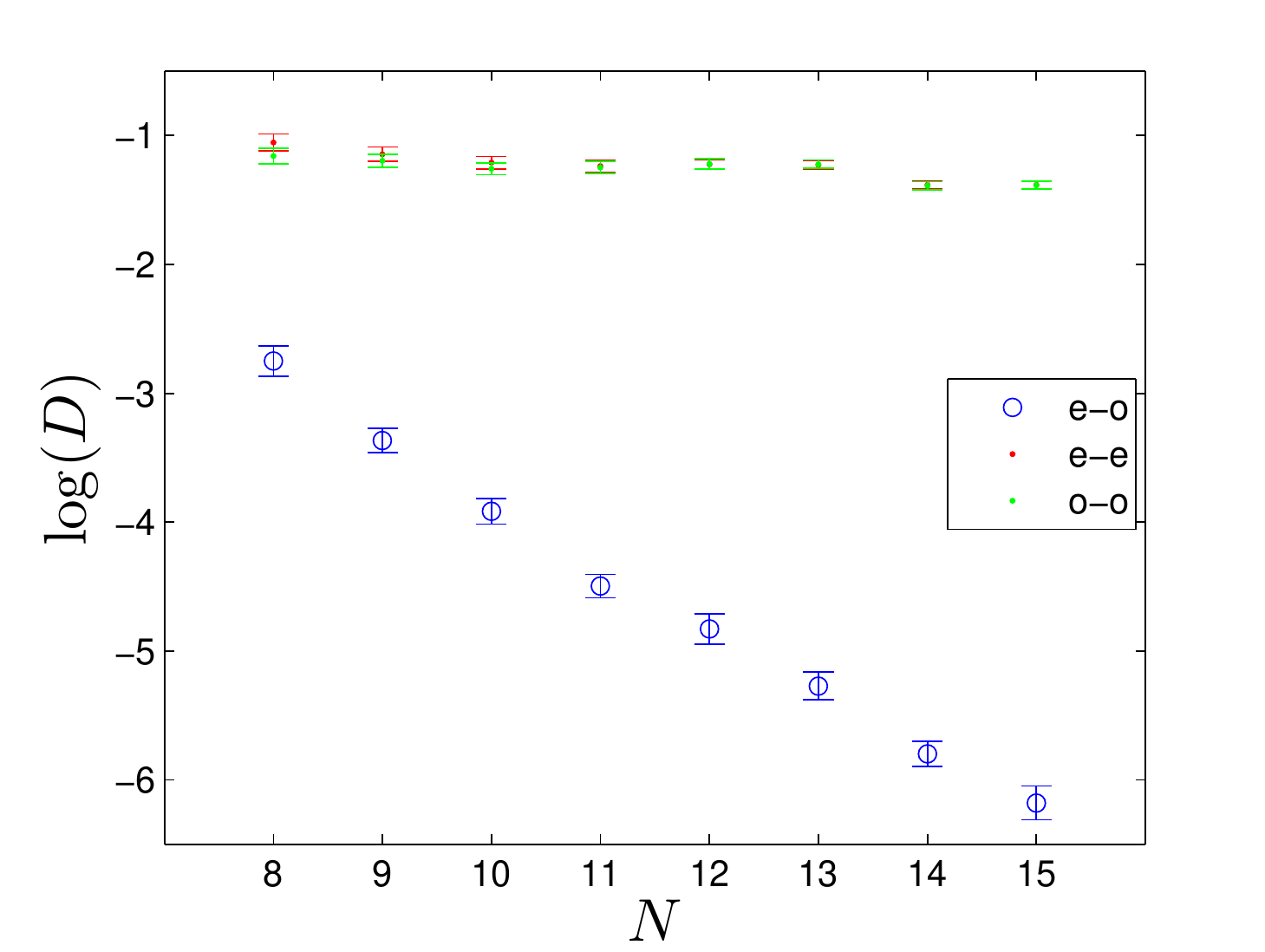}
    \end{minipage}}
  \subfigure[\, Eigenstate distance in phase F3]{
    \label{fig:F1} 
    \begin{minipage}[b]{\columnwidth}
      \centering
      \includegraphics[width = \textwidth]{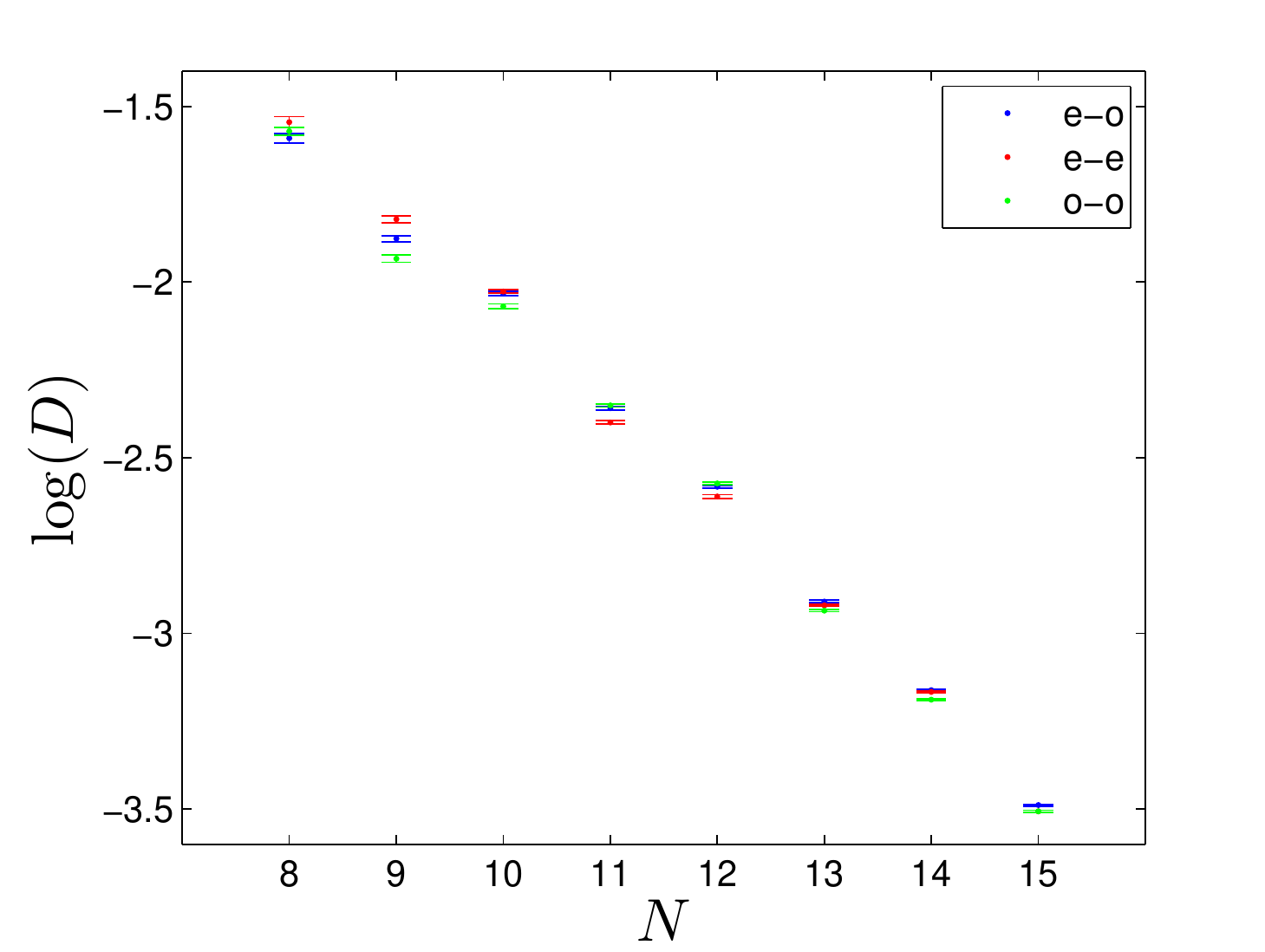}
    \end{minipage}}
  \caption{(color online)  Averages of $\log{(D)}$ in phases F1 and F3, respectively, where $D$ is the `eigenstate distance' defined in the text.. The energy density range we used
in F1 is from the first excited state in each sector up to $u_c - 0.02$ where $u_c$ is the energy density at the phase boundary between F1 and F2, whereas in F3 we used the phase's full energy density range. $N$ is the total number of spins varying from 8 to 15.
The exponential decrease of $D$ with increasing $N$ indicates thermalization.
The error bars come from averaging over 100 realizations.}
 \label{fig:eigenstate_distance} 
\end{figure}

Next, we examine a `distance' between two eigenstates that are adjacent in the energy spectrum by comparing their probability distributions for the total spin $S_z$.  We define this distance between eigenstate 1 and eigenstate 2 as
\begin{equation}
D_{12} = \sum_{S_z=-N/2}^{N/2}\big|P_1(S_z)-P_2(S_z)\big| ~,
\end{equation}
where $P_1(S_z)$ and $P_2(S_z)$ are the probability distributions of $S_z$ in eigenstates 1 and 2, respectively.  We tested three different
distances: $D_{eo}$ the distance between an even parity state (eigenstate 1) and the nearest-energy odd parity state (eigenstate 2),
and similarly for $D_{ee}$ and $D_{oo}$.  If a system thermalizes, each eigenstate is equivalent to a microcanonical ensemble characterized by its energy.  For two eigenstates that are adjacent in energy, the energy difference $\Delta{}E\sim{}2^{-N}$, therefore we expect the eigenstate distances $D_{eo}$, $D_{ee}$ and $D_{oo}$ should decrease exponentially with $N$.  In phase F1, since the spectrum consists of nearly-degenerate pairs of states, the energy differences satisfy $\Delta{}E_{eo}\ll\Delta{}E_{ee}\,,\,\Delta{}E_{oo}$.
Thus we expect in phase F1, $D_{eo}\ll{}D_{ee}\,,\,D_{oo}$.
In addition, if we choose the upper bound of the energy window we average over to be well within the F1 phase, we would also expect that Eq. (11) holds, so the exponential decay rate of $D_{eo}$ would be greater than those of $D_{ee}$ and $D_{oo}$.
Meanwhile, in phases other than F1, we expect all three $D$'s are well coincident. The numerical results are shown in Fig. 3.  As we expected, all these eigenstate distances decay exponentially with $N$.  In addition, $D_{eo}$ in phase F1 is much smaller and decreasing much faster than the other two distances, and in phase F3 all three $D$'s are well coincident.  This demonstrates the clear distinction between phases F1 and F3, and further tests the thermalization in phase F3.

\begin{figure}[!htbp]
\begin{center}
\includegraphics[width = \columnwidth]{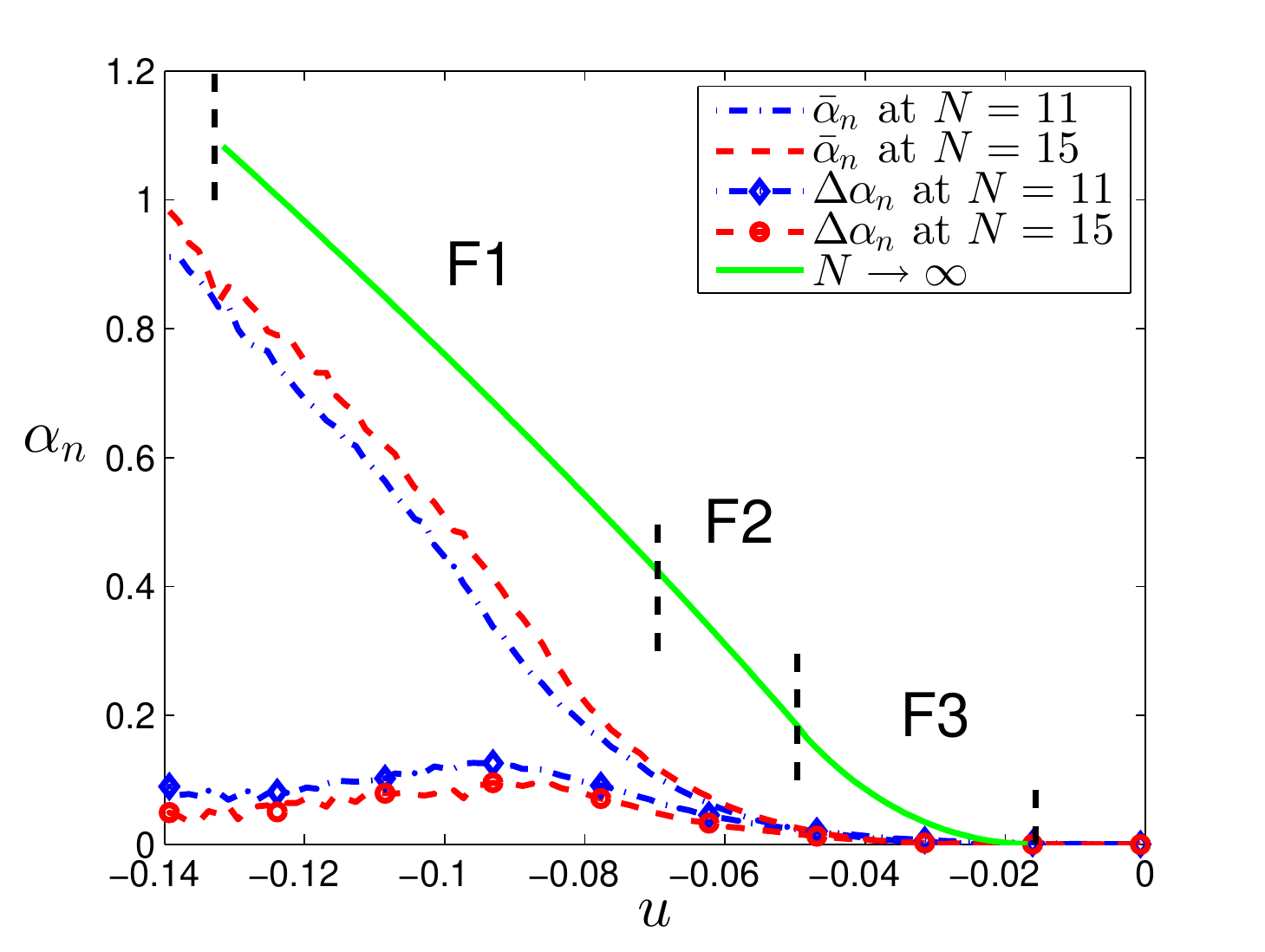}
\caption{(color online) The mean $\bar{\alpha}_n$ and the standard deviation $\Delta\alpha_n$ of the quantity $\alpha_n$ defined in Eq. (16).  The number of realizations is 1600 for $N=11$ (blue dash-dotted lines) and 100 for $N=15$ (red dashed lines).  The solid (green) line gives the theoretical quantity $\alpha(u,\Gamma)$ defined in Eq. (9) for the system size $N\rightarrow\infty$.}
\label{fig:mean}
\end{center}
\end{figure}


In the ferromagnetic phases, the system spontaneously flips between
magnetization up and down at a rate that behaves as $\sim\exp(-N\alpha(u,\Gamma))$, where the quantity $\alpha(u,\Gamma)$ is defined in Eq. (9).
The probability of the system having total magnetization zero (or 1/2 for systems with odd $N$) also behaves as $\sim\exp(-N\alpha(u,\Gamma))$.
Thus from a single many-body eigenstate $|n\rangle$ we can obtain an estimate of the quantity $\alpha$ as
\begin{equation}
\alpha_n = \frac{1}{N}\ln\frac{\max_{S_z}\{P_n(S_z)\}}{P_n(S_z=0\textrm{ or }1/2)}  ~,
\end{equation}
where $P_n(S_z)$ is the probability distribution of the total magnetization $S_z$ in this eigenstate.  If the ETH is true, the magnetization thermalizes and these estimates $\alpha_n$ will converge to $\alpha(u,\Gamma)$ in the limit of large $N$.

In Fig. 4 we show the mean ($\bar{\alpha}_n$) and the standard deviation ($\Delta\alpha_n$) of $\alpha_n$ within energy bins for systems of size $N=11$ and $N=15$.  The standard deviation decreases with increasing $N$, as expected for these Schr\"odinger cat states that obey the ETH.  In the limit of large $N$ every eigenstate at a given energy density will have the same probability distribution of the total magnetization.  That distribution is a thermal distribution for the magnetizations that are accessed by thermal excitation and `tails' due to quantum tunneling in the remainder of the distribution.

The model we are studying here is a ferromagnet ($H_0$) with a small added spin-glass term ($H_1$) in its Hamiltonian.  In terms of its affect on the system's thermodynamics, the relative strength of the spin-glass term scales as $\sim N^{-1/4}$ for the case $p=3/4$ that we have here.  Thus it is perhaps reasonable to expect the mean value of $\alpha_n$ to exhibit a finite-size correction that vanishes for large $N$ as $\sim N^{-1/4}$.  The spin-glass term weakly frustrates the ferromagnetism, so will cause a reduction in the apparent $\alpha$.  Given the very modest range of $N$ for which we can do exact diagonalizations, this $\sim N^{-1/4}$ is a very slow convergence towards the thermodynamic limit.  In Fig. 4 we also show the expected value of $\alpha$ in the limit of large $N$.  As $N$ is increased from 11 to 15, $\bar{\alpha}_n$ does indeed increase slowly toward $\alpha(u,\Gamma)$, as expected.

Thus the results of these exact diagonalizations provide some numerical support for the theoretical results derived above assuming this system obeys the ETH.  Of course, given the very small size systems that can be diagonalized, there are strong finite-size effects that make a detailed demonstration of thermalization in the large system limit not possible.  This is the situation with all numerical tests of thermalization.

In addition to the model discussed above, we tested two other ways of adding disorder to $H_0$.  The other Hamiltonians that we diagonalized are
\begin{equation}
H' = H_0 + \frac{1}{N^{p'}}\sum_{i=1}^{N}\varepsilon_{i}s_{i}^{x} ~,
\end{equation}
and
\begin{equation}
H'' = -\frac{1}{N}\sum_{1=i<j}^N (\eta_i+1)(\eta_j+1)s_{i}^{z}s_{j}^{z} - \Gamma\sum_{i=1}^N s_{i}^{x} ~,
\end{equation}
where in $H'$, the $\varepsilon_{i}$ are also independent Gaussian random variables of mean zero and variance one, and the power $p'$ here satisfies $0<p'<1/2$; specifically we looked at $p'=1/4$.  In $H''$, the $\eta_i$ are independent random variables uniformly distributed in the interval $[-1,1]$.  These latter two models only have $N$ random parameters, one per spin, unlike (2) which has one per pair of spins: $N(N-1)/2$ random couplings.  This difference appears to be quantitatively important, as neither of these latter two models showed good evidence of thermalization under the tests illustrated above for the sizes that can be exactly diagonalized.  But we expect that in the limit of large $N$ this difference should go away, with all of these models thermalizing.



\section{Acknowledgements}

We thank Mike Kolodrubetz (MK) and Hyungwon Kim for discussions, including preliminary numerical work by MK on a related model.  This work was supported in part by the National Science Foundation under DMR0819860, and by the DARPA OLE program.

\end{document}